\documentclass[utf8]{frontiersSCNS} 

\usepackage{url,hyperref,microtype,subcaption}
\usepackage[onehalfspacing]{setspace}

\usepackage[T1]{fontenc}
\usepackage[utf8]{inputenc}
\usepackage{color}
\usepackage{array}
\usepackage{float}
\usepackage{url}
\usepackage{amsmath}
\usepackage{amssymb}
\usepackage{graphicx}
\usepackage{wasysym}
\usepackage{blindtext}
\PassOptionsToPackage{hyphens}{url}
\usepackage{breakurl}
\usepackage[linesnumbered,ruled]{algorithm2e}

\def\keyFont{\fontsize{8}{11}\helveticabold }
\def\firstAuthorLast{Sample {et~al.}} 
\def\Authors{Gang Yao\,$^{1}$, Ashwin Dani\,$^{1,*}$ }
\def\Address{$^{1}$Department of Electrical and Computer Engineering, University of Connecticut, Storrs, CT, 06269, USA }

\def\corrAuthor{Corresponding Author}

\def\corrEmail{ashwin.dani@uconn.edu}

\begin{document}
\onecolumn
\firstpage{1}

\title[Running Title]{Visual Tracking Using Sparse Coding and Earth Mover's Distance} 

\author[\firstAuthorLast ]{\Authors} 
\address{} 
\correspondance{} 

\extraAuth{}

\maketitle

\begin{abstract}

\section{}
An efficient iterative Earth Mover's Distance (iEMD) algorithm for visual tracking is proposed in this paper. The Earth Mover's Distance (EMD) is used as the similarity measure to search for the optimal template candidates in feature-spatial space in a video sequence. The computation of the EMD is formulated as the transportation problem from linear programming. The efficiency of the EMD optimization problem limits its use for visual tracking. To alleviate this problem, a transportation-simplex method is used for EMD optimization and a monotonically convergent iterative optimization algorithm is developed. The local sparse representation is used as the appearance models for the iEMD tracker. The maximum-alignment-pooling method is used for constructing a sparse coding histogram which reduces the computational complexity of the EMD optimization. The template update algorithm based on the EMD is also presented. The iEMD tracking algorithm assumes small inter-frame movement in order to guarantee convergence. When the camera is mounted on a moving robot, e.g., a flying quadcopter, the camera could experience a sudden and rapid motion leading to large inter-frame movements. To ensure that the tracking algorithm converges, a gyro-aided extension of the iEMD tracker is presented, where synchronized gyroscope information is utilized to compensate for the rotation of the camera. The iEMD algorithm's performance is evaluated using eight publicly available datasets. The performance of the iEMD algorithm is compared with seven state-of-the-art tracking algorithms based on relative percentage overlap. The robustness of this algorithm for large inter-frame displacements is also illustrated. 

\tiny
 \keyFont{ \section{Keywords:} Visual tracking; Earth Mover's Distance; Sparse coding; Gyro-aided tracking; Transportation Simplex Method} 
\end{abstract}

\section{Introduction}

Visual tracking is an important problem in the field of computer vision.
Given a sequence of images, tracking is the procedure of generating
the inference about the motion of the target. There are a variety
of applications for visual tracking. The information generated from
these images by the tracking algorithm can be
utilized by vehicle navigation, human-robot interaction, and motion-based
recognition algorithms \citep{dani2013,RavichandarIntentionIMM,chwa2016range}.
Visual tracking algorithms provide important information
for visual simultaneous localization and mapping (SLAM), structure
from motion (SfM) and video-based control \citep{dani2012single,Yang2015,davison2007monoslam}.

Image-based tracking algorithms are categorized as point tracking,
kernel tracking, or silhouette tracking \citep{yilmaz2006object}.
Distinguishing features, such as color, shape, and region are selected
to identify objects for visual tracking. Modeling the object which
could adapt to the slowly changing appearance is challenging, due
to the illumination variants, object deformation, occlusion, motion
blur or background clutters. Supervised or unsupervised online learning
algorithms are often used to robustly find and update the \textcolor{black}{distinguishing}
features of the object, such as using variance ratios of the feature
value's log likelihood \citep{collins2005online}, the online Ada-boost
feature selection method \citep{grabner2006line} and incremental learning
\citep{ross2008incremental}.

Approaches in visual tracking could be generally classified into two
groups, either generative methods or discriminative methods. For generative
methods, the tracked object is modeled based on the selected features,
such as the color histogram, sparse coding representation or kernels.
Then, correspondence or similarity measurement between the target
and the candidate across frames is constructed. Similarity measurements
are derived through several methods, such as the Normalized Cross
Correlation (NCC) \citep{bolme2010visual,zhu2016clustering}, the Earth
Mover's Distance (EMD) \citep{zhao2010differential,oron2012locally,karavasilis2011visual},
the Bhattacharyya Coefficient (BC) \citep{comaniciu2003kernel} and
point-to-set distance metric \citep{wang2015affine,wang2016adaptive}.
Location of the candidate object in the consecutive frames is estimated
by using the Kalman filter, particle filter or gradient descent method.
Discriminative methods regard tracking as a classification problem
and build a classifier or ensemble of classifiers to distinguish the
object from the background. Representative classification tracking
algorithms are the structured Support Vector Machine (SVM) \citep{hare2011struck},
Convolutional Neural Nets \citep{li2016convolutional}. Ensemble based
algorithms such as ensemble tracking \citep{avidan2007ensemble}, multiple
instance learning (MIL) \citep{babenko2011robust}, \textcolor{black}{and
o}nline boosting tracker \citep{grabner2006line}.

In order to robustly track moving objects in challenging situations,
many tracking frameworks are proposed. Tracking algorithms with Bayesian
filtering are developed to track moving objects. These algorithms
can handle complete occlusion \citep{zivkovic2009approximate}. The
non-adaptive methods, usually only model the object from the first
frame. Although less error prone to occlusions and drift, they are
hard to track the object undergoing appearance variations. However,
adaptive meth\textcolor{black}{ods are u}sually prone to drift because
they rely on self updates of an online learning method. In order to
deal with this problem, combining adaptive methods with the complementary
tracking approaches leads to more stable results. For example, parallel
robust online simple tracking (PROST) framework combines three different
trackers \citep{santner2010prost}: tracking-learning-detection (TLD)
framework uses P-N experts to make the decision on the location of
the moving object, based on the results from the Median-Flow tracker
and detectors \citep{kalal2012tracking}, and online adaptive hidden
Markov model for multi-tracker fusion \citep{vojir2016online}. 

The emphasis\textcolor{black}{{} of thi}s paper is on the similarity
measurement and target localization. The EMD is adopted as the similarity
measure and an efficient \textcolor{black}{iterative} EMD algorithm
is proposed for visual tracking. The contributions of the paper are
summarized as follows: 
\begin{itemize}
\item \textcolor{black}{The maximum-alignment-pooling method for local sparse
coding is used to build a histogram of appearance model. An iEMD tracking
algorithm is developed based on this local sparse coding representation
of the appearance model. It is shown using videos from publicly available
benchmark datasets that the iEMD tracker shows good performance in
terms of percentage overlap compared to the state-of-the-art trackers
available in literature.}
\item Gyro-measurements are used to compensate for the pan, tilt, and roll
of the camera. Then the iEMD visual tracking algorithm is used to
track the target after compensating for the movement of the camera.
By this method, the convergence of the algorithm is ensured, thus
providing a more robust tracker which is more capable of real-world
tracking tasks. 
\end{itemize}
The paper is organized as follows. Related work on the computation
of the EMD and its application for visual tracking is illustrated
in Section 2. In Section 3, the iEMD algorithm for visual tracking
is developed. In Section 4, the target is modeled as the sparse coding
histogram. For the sparse coding histogram, the maximum-alignment-pooling
method is proposed to represent the local image patches. In Section
5, two extensions of the iEMD algorithm that includes the template
update method, and the method of using the gyroscope data for ego-motion
compensation are discussed. In Section 6, the iEMD tracker is validated
on eight publicly available datasets, and the comparisons with seven
state-of-the-art trackers are shown. Experimental results using the
gyro-aided iEMD algorithm are compared with tracking results without
gyroscope information. The conclusions are given in Section 7.
\section{Related Work}

In real-world tracking applications, variations in appearance are
a common phenomenon caused by illumination changes, moderate pose
changes or partial occlusions. The Earth Mover's Distance (EMD) as
a similarity measure, also known as 1-Wasserstein distance \citep{Baum2015,guerriero2010shooting},
is robust to these situations \citep{rubner2000earth}. However, the
major problem with the EMD is its computational complexity. Several
algorithms for the efficient computation of the EMD are proposed.
For example, the EMD-$L_{1}$ algorithm is used for histogram comparison
\citep{ling2007efficient} and the EMDs are computed with the thresholded
ground distances \citep{pele2009fast}. In the context of visual tracking,
although the EMD has the merit of being robust to moderate appearance
variations, the efficiency of the computation is still a problem.
\textcolor{black}{Since solving the EMD is a transportation problem
– a linear programming problem \citep{rubner2000earth},} the direct
differential method cannot be used. There are some efforts to employ
the EMD for object tracking. The Differential Earth Mover's Distance
(DEMD) algorithm \citep{zhao2010differential} is first proposed for
visual tracking, which adopts the sensitivity analysis to approximate
the derivative of the EMD. However, the\textcolor{black}{{} selection
of the basic variables }and the process of identifying and deleting
the redundant constraints still affect the efficiency of the algorithm
\citep{zhao2010differential}. The DEMD algorithm combined with the
Gaussian Mixture Model (GMM), which has fewer parameters for EMD optimization,
is proposed in \citep{karavasilis2011visual}. The EMD as the similarity
measure combined with the particle filter for visual tracking is proposed
in \citep{oron2012locally}.

Sparse coding has been successfully applied to visual
tracking \citep{zhang2013sparse}. In sparse coding for visual tracking,
the largest sum of the sparse coefficients or the smallest reconstruction
error is used as the metric to find the target from the candidate
templates using particle filter \citep{mei2009robust,jia2016visual}.
The sparse coding process is usually the $L_{1}$ norm minimization
problem, which makes the sparse representation and dictionary learning
computationally expensive. To reduce the computational complexity,
the sparse representation as the appearance model is combined with
the Mean-shift \citep{liu2011robust} or Mean-transform method \citep{zhang2014pyramid}.
After a small number of iterations by these methods, the maximum value
of the Bhattacharyya coefficient corresponding to the best candidate
is obtained. 

The success of the gradient descent based tracking algorithm depends
on the assumption that the object motion is smooth and contains only
small displacements \citep{yilmaz2006object}. However, in practice,
this assumption is always violated due to the abrupt rotation and
shaking movement of the camera mounted on a robot, such as a flying
quadcopter. Efforts have been made to combine the gyroscope data with
tracking algorithms, such as the Kanade-Lucas-Tomasi (KLT) tracker
or the MI tracker \citep{Hwangbo2011,ravichandar2014gyro,park2013novel}.
To robustly track a static object using a moving camera, gyroscope
data are directly utilized to estimate the initial location of the
static object. When both the camera and the object being tracked are
in motion, the gyroscope sensor data are utilized to compensate for
the rotation of the camera, because rotation has a greater impact
on the positional changes compared with the translation in video frames.
Then, the visual tracking algorithm is applied to track the moving
object. The robustness of the tracking algorithm is improved due to
the compensation of the camera's ego-motion. Therefore, our method
makes the EMD tracker more robust to this situation.

\section{Iterative EMD Tracking Algorithm}

In the context of visual tracking, first a feature space is chosen
to characterize the object, then, the target model and the candidate
model are built in the feature-spatial space. The probability density
functions (histograms) representing the target model and the candidate
model are \citep{comaniciu2003kernel}

target model: $\hat{\mathbf{p}}=\left\{ \hat{p}_{u}\right\} _{u=1,\ldots,N_{T}}$
and $\sum_{1}^{N_{T}}\hat{p}_{u}=1$

candidate model: $\hat{\mathbf{q}}(\mathbf{y})=\left\{ \hat{q}_{v}(\mathbf{y})\right\} _{u=1,\ldots,N_{C}}$
and $\sum_{1}^{N_{C}}\hat{q}_{v}(\mathbf{y})=1$, where $\hat{p}_{u}$
is the weight of the $u\mathrm{th}$ bin of the target model $\hat{\mathbf{p}}$,
assuming the center of the template target is at $\left(0,0\right)$,
$\hat{q}_{v}$ is the weight of the $v\mathrm{th}$ bin of the candidate
model $\hat{\mathbf{q}}(\mathbf{y})$, assuming the center of the
template candidate is at $\mathbf{y}$, $N_{T}$ and $N_{C}$ are
the numbers of the bins.

Based on the target model and the candidate model, the dissimilarity
function is denoted as $f(\hat{\mathbf{p}},\hat{\mathbf{q}}(\mathbf{y}))$.
The optimization problem for tracking is to estimate the optimal displacement
$\hat{\mathbf{y}}$ which gives the smallest value of $f(\hat{\mathbf{p}},\hat{\mathbf{q}}(\mathbf{y}))$.
Thus, the optimization problem is formulated as 
\begin{equation}
\hat{\mathbf{y}}\,=\arg\ \underset{\mathbf{y}}{\min}\ f(\hat{\mathbf{p}},\hat{\mathbf{q}}(\mathbf{y}))\label{eq:1}
\end{equation}
In (\ref{eq:1}), the center of the template target is assumed to
be positioned at $\left(0,0\right)$, and the center of the template
candidate is at $\mathbf{y}$. The goal is to find the candidate model
located at $\hat{\mathbf{y}}$ that gives the smallest value of the
dissimilarity function $f(\hat{\mathbf{p}},\hat{\mathbf{q}}(\mathbf{y}))$.
The differential tracking approaches are usually applied to solve
this optimization problem, with the assumption that the displacement
of the target between two consecutive frames is very small.

The optimization problem in (\ref{eq:1}) is solved using the iEMD
algorithm as described in the following sub-sections. The iEMD algorithm
iterates between finding the smallest EMD between template target
and the template candidate based on the current position $\mathbf{y}_{k}$
by the transportation-simplex method (see Section \ref{sub:Iterative-EMD-Tracking}
for details) and finding the best position $\mathbf{y}_{k+1}$ leading
to the smallest EMD by gradient method (see Section \ref{sub:Gradient-Method-to}
for details).
\subsection{EMD as a Similarity Measure}

\begin{figure}[H]
\begin{centering}
\includegraphics[width=0.95\columnwidth]{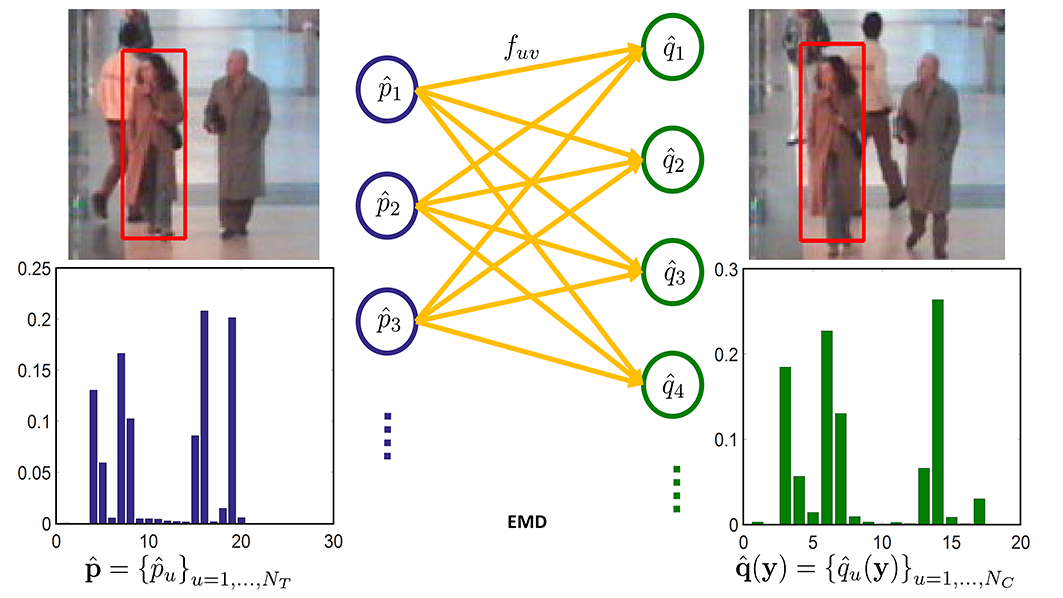}\protect\caption{EMD comparison of the two templates.\label{fig:EMD-comparison-of}}
\par\end{centering}
\end{figure}

In this section, the Earth Mover's Distance (EMD) between the target
model $\hat{\mathbf{p}}$ and the candidate model $\hat{\mathbf{q}}(\mathbf{y})$
is used as the similarity measure. Solving the EMD is a transportation
problem \textcolor{black}{–} a linear programming problem as shown
in Fig. \ref{fig:EMD-comparison-of}. Intuitively, given the target
model and the candidate model, one is thought of as a set of factories
and the other as a set of shops. Suppose that a given amount of goods
produced by the factories are required to be delivered to the shops,
each with a given limited capacity. The cost to ship a unit of goods
from every factory to different shops is not equivalent. Then the
EMD is considered as the smallest overall cost of sending the weights
(goods) from the target model to the candidate model. The EMD is defined
as \citep{rubner2000earth} 
\begin{equation}
D^{\star}(f_{uv}(\mathbf{\hat{\mathbf{p}},\hat{\mathbf{q}}}(\mathbf{y})))\triangleq\underset{f_{uv}}{\min}\,\left(\sum_{u=1}^{N_{T}}\sum_{v=1}^{N_{C}}d_{uv}f_{uv}(\hat{\mathbf{p}},\hat{\mathbf{q}}(\mathbf{y}))\right)\label{eq:4-1}
\end{equation}
subject to 
\begin{equation}
\sum_{u=1}^{N_{T}}f_{uv}(\hat{\mathbf{p}},\hat{\mathbf{q}}(\mathbf{y}))=w_{C,v},1\leq v\leq N_{C}\label{eq:5-1}
\end{equation}

\begin{equation}
\sum_{v=1}^{N_{C}}f_{uv}(\hat{\mathbf{p}},\hat{\mathbf{q}}(\mathbf{y}))=w_{T,u},1\leq u\leq N_{T}\label{eq:6-1}
\end{equation}

\begin{equation}
\sum_{u=1}^{N_{T}}\sum_{v=1}^{N_{C}}f_{uv}(\hat{\mathbf{p}},\hat{\mathbf{q}}(\mathbf{y}))=1\label{eq:7-1}
\end{equation}
\begin{equation}
f_{uv}(\hat{\mathbf{p}},\hat{\mathbf{q}}(\mathbf{y}))\geq0,1\leq u\leq N_{T},1\leq v\leq N_{C}\label{eq:8-1}
\end{equation}
where $D^{\star}$ is the optimal solution to this transportation
problem, $f_{uv}(\hat{\mathbf{p}},\hat{\mathbf{q}}(\mathbf{y}))$
is the flow (weight) from the $u\mathrm{th}$ bin of $\hat{\mathbf{p}}$
to the $v\mathrm{th}$ bin of $\hat{\mathbf{q}}(\mathbf{y})$, $d_{uv}$
is the ground distance (cost) between the $u\mathrm{th}$ and the
$v\mathrm{th}$ bins, the subscript $T$ denotes the object target
and $C$ is the object candidate, $w_{T,u}$ is the weight from the
$u\mathrm{th}$ bin of $\hat{\mathbf{p}}$, and $w_{C,v}$ is the
weight from the $v\mathrm{th}$ bin of $\hat{\mathbf{q}}(\mathbf{y})$. 

\subsection{EMD as a Function of Weights\label{sub:Iterative-EMD-Tracking}}

Writing the above equation set (\ref{eq:4-1})-(\ref{eq:8-1}) in
a matrix form as 
\begin{equation}
\begin{array}{cc}
 & D^{\star}=\underset{\mathbf{f}}{\mathrm{min\,}}\mathbf{d^{\mathit{T}}}\mathbf{f}\\
\mathrm{s.t.} & \mathbf{H}\mathbf{f}=\mathbf{w};\,\mathbf{f}>0
\end{array}\label{eq:7}
\end{equation}
where the $\mathbf{d}=[d_{11},\cdots d_{1N_{C}},\cdots d_{N_{T}1},\cdots d_{N_{T}N_{C}}]^{T}\in\mathbb{R}^{N_{T}N_{C}}$
is the ground distance vector, $\mathbf{f}=[f_{11},\cdots f_{1,N_{C}},\cdots f_{N_{T}1,}\cdots f_{N_{T}N_{C}}]^{T}\in\mathbb{R}^{N_{T}N_{C}}$
is the flow vector, $\mathbf{w}=[\mathbf{w_{\mathit{N_{C}}}^{\mathit{T}}},\mathbf{w}_{N_{T}}^{T}]^{T}\in\mathbb{R}^{N_{T}+N_{C}}$
is the weight vector consisting of the weight vectors $\mathbf{w_{\mathit{N_{C}}}}\in\mathbb{R}^{N_{C}}$
from $\hat{\mathbf{q}}(\mathbf{y})$ and $\mathbf{w_{\mathit{N_{T}}}}\in\mathbb{R}^{N_{T}}$
from $\hat{\mathbf{p}}$, and $\mathbf{H\in\mathbb{R}}^{(N_{T}+N_{C})\times N_{T}N_{C}}$
is the matrix which consists of $0s$ and $1s$.

In order to relate the EMD with the weight vector, the above primal
problem in (\ref{eq:7}) is restated in its dual form as \citep{dantzig2006linear}
\begin{equation}
\begin{array}{cc}
 & D^{\star}=\underset{\mathbf{\pi}}{\mathrm{max}\,}\mathbf{w^{\mathit{T}}}\mathbf{\mathbf{\pi}}\\
\mathrm{s.t.} & \mathbf{H}^{T}\mathbf{\pi}\leq\mathbf{d}
\end{array}\label{eq:8}
\end{equation}
where $\pi\in\mathbb{R}^{N_{T}+N_{C}}$ is a vector of variables to
be optimized in the dual problem. By solving this dual problem in
(\ref{eq:8}), the optimal solution $D^{\star}$ is calculated and
directly represented as the linear equation of weights. However, considering
the computation efficiency, the optimal solution (EMD) is first calculated
from the primal problem in (\ref{eq:7}) using the transportation-simplex
method, and then the EMD is represented as the function of the weights
by the matrix transformation.

Using the transportation-simplex method \citep{rubner2000earth}, the
optimal solution to the EMD problem in (\ref{eq:7}) is calculated.
The transportation-simplex method is a streamlined simplex algorithm,
which is built on the special structure of the transportation problem.
In order to reduce the number of iterations of the transportation-simplex
method, the Russell's method is used to compute the initial basic
feasible solution \citep{rubner2000earth,ling2007efficient}. The DEMD
algorithm \citep{zhao2010differential} uses the standard simplex method
to compute the optimal solution to the linear optimization problem
in (\ref{eq:7}). Compared with the standard simplex method, the transportation-simplex
method greatly decreases the number of operations \citep{ling2007efficient}.
Thus, the iEMD algorithm is more efficient in terms of the number
of operations to solve the EMD problem compared with the DEMD algorithm
in \citep{zhao2010differential}.

The computation of the EMD is a transportation problem, which has
exactly $N_{T}+N_{C}-1$ basic variables $\mathbf{f_{\mathit{B}}}\in\mathbb{R}^{N_{T}+N_{C}-1}$,
and each constraint is a linear combination of the other $N_{T}+N_{C}-1$
constraints, which could be considered as redundant and discarded
\citep{dantzig2006linear}. Based on the optimal solution to the linear
programming problem, the flow vector is separated into basic variables
and non-basic variables as $\mathbf{f}=[\mathbf{f_{\mathit{B}}^{\mathit{T}},\mathbf{\mathbf{f_{\mathit{NB}}^{\mathit{T}}}}}]^{T}\in\mathbb{R}^{N_{T}N_{C}}$,
and the ground distance vector $\mathbf{d}$ and $\mathbf{H}$ will
be transformed as $\mathbf{d}=[\mathbf{d_{\mathit{B}}^{\mathit{T}},\mathbf{\mathbf{d_{\mathit{NB}}^{\mathit{T}}}}}]^{T}\in\mathbb{R}^{N_{T}N_{C}}$
and $\mathbf{H}=[\mathbf{H_{\mathit{B}},\mathbf{\mathbf{H_{\mathit{NB}}}}}]^{T}\mathbf{\in\mathbb{R}}^{(N_{T}+N_{C})\times N_{T}N_{C}}$,
where $\mathbf{d_{\mathit{B}}}\in\mathbb{R}^{N_{T}+N_{C}-1}$, and
$\mathbf{H_{\mathit{B}}}\in\mathbb{R}^{(N_{T}+N_{C})\times(N_{T}+N_{C}-1)}$.
In order to derive the EMD as a function of the weights of the candidate
model, the matrix transformation is performed. First, the last row
of the constraint matrices (\ref{eq:7}) is deleted which is considered
as the redundant constraint and then the matrices $\mathbf{H_{\mathit{B}}}$,
$\mathbf{H}$, and $\mathbf{w}$ are formulated as $\mathbf{H_{\mathit{B}}^{\ast}}\in\mathbb{R}^{(N_{T}+N_{C}-1)\times(N_{T}+N_{C}-1)}$,
$\mathbf{H^{\ast}}=[\mathbf{H_{\mathit{B}}^{\ast},\mathbf{\mathbf{H_{\mathit{NB}}^{\ast}}}}]^{T}\mathbf{\in\mathbb{R}}^{(N_{T}+N_{C}-1)\times N_{T}N_{C}}$
and $\mathbf{w^{\ast}}=[\mathbf{w_{\mathit{N_{C}}}^{\mathit{T}}},\mathbf{w}_{N_{T}-1}^{T}]^{T}\in\mathbb{R}^{N_{T}+N_{C}-1}$.

The problem in (\ref{eq:7}) is reformulated based on the optimal
solution as
\begin{align}
D^{\star}-\mathbf{d_{B}}^{T}\mathbf{f_{B}}-\mathbf{d_{NB}}^{T}\mathbf{f_{NB}} & =0\label{eq:9-2}\\
\mathbf{H_{B}^{\ast}}\mathbf{f_{B}}+\mathbf{H_{NB}^{\ast}}\mathbf{f_{NB}} & =\mathbf{w^{\ast}}\label{eq:10-2}
\end{align}
Left multiplying (\ref{eq:10-2}) with $\mathbf{H_{B}^{\ast-1}}$
yields 
\begin{equation}
\mathbf{f_{B}}+\mathbf{H_{B}^{\ast-1}\mathbf{H_{NB}^{\ast}}}\mathbf{f_{NB}}=\mathbf{H_{B}^{\ast-1}}\mathbf{w}^{\ast}\label{eq:10-1}
\end{equation}
Left multiplying (\ref{eq:10-1}) by $\mathbf{d_{B}}^{T}$ and adding
the resultant to (\ref{eq:9-2}) gives 
\begin{equation}
D^{\star}+(-\mathbf{d_{NB}}^{T}+\mathbf{M\mathbf{H_{NB}^{\ast}}})\mathbf{f_{NB}}=\mathbf{M}\mathbf{w}^{\ast}\label{eq:12}
\end{equation}
where $\mathbf{M}=\mathbf{d_{B}}^{\mathit{T}}\mathbf{H_{B}^{\ast-1}}$
is a $N_{C}+N_{T}-1$-dimensional vector. Since $\mathbf{f_{NB}}=\mathbf{0}_{N_{T}N_{C}-N_{T}-N_{C}+1}$,
using (\ref{eq:12}) the EMD $D^{\star}$ is given by 
\begin{equation}
D^{\star}=\mathbf{M[\mathbf{w_{\mathit{N_{C}}}^{\mathit{T}}},0]_{\mathit{N_{C}+N_{T}-1}}^{\mathit{T}}+M[0,\mathbf{w_{\mathit{N_{T}-1}}^{\mathit{T}}}]_{\mathit{N_{C}+N_{T}-1}}^{\mathit{T}}}\label{eq:9}
\end{equation}

\subsection{Gradient Method to Find the Template Displacement\label{sub:Gradient-Method-to}}

Based on the equation (\ref{eq:9}) , the gradient method is utilized
to find the displacement $\mathbf{y}$ of the target candidate as
\begin{equation}
\frac{\partial D^{\star}}{\partial\mathbf{y}}=\mathbf{M}\left[\mathbf{\frac{\partial w_{\mathit{N_{C}}}^{\mathit{T}}}{\partial y}},0\right]_{N_{C}+N_{T}-1}^{T}\label{eq:10}
\end{equation}
The optimal location $\hat{\mathbf{y}}$ of the template candidate
$\hat{\mathbf{q}}(\mathbf{y})$ is found by iteratively performing:
(1) calculate the smallest EMD and reformulate it as (\ref{eq:9});
(2) search for the new location of the template candidate along the
direction of (\ref{eq:10}). When the EMD no longer decreases, the
iteration stops. By this method, the best match of the template target
and the template candidate will be found. The EMD plays three roles
in this algorithm: (1) it provides a metric of the matching between
the template target and the template candidate; (2) it assigns more
weights to the best matches between the histogram bins and assigns
smaller weights or no weights to unmatched bins by linear optimization;
(3) matched bins are used for finding the location of the template
candidate, and the gradient vector of the EMD for searching the optimal
displacement is calculated. 

The pseudo-code for the iEMD tracking algorithm is given in Algorithm
1.

\begin{algorithm}[h]
Set the maximum iteration number ${n_{iter}}$\;

Calculate the target model from the image ${I_{0}}$ using (\ref{eq:sparse target model})\;

Get the new image frame $I_{k+1}$\;

Construct the candidate model from $I_{k+1}$ using (\ref{eq:sparse candidate model})\;

Compute $EMD_{pre}$ between the target model and the candidate model\;

\For{$n$=0 to $n_{iter}$}{

Represent the $EMD_{pre}$ by its weight vector $\mathbf{w}^{T}$
using (\ref{eq:9})\;

Calculate the derivative of the $EMD$ with respect to the displacement
$\mathbf{y}$ using (\ref{eq:10})\;

Move the template candidate in $I_{k+1}$ along the gradient vector
by one pixel\;

Compute $EMD$ between the target model and the new candidate model\;

\If{$EMD_{pre}<EMD$ }{ break\; }

\Else{ n=n+1\; Set the $EMD_{pre}=EMD$ \; } } \protect

\protect\caption{iEMD tracking algorithm}

\label{Iterative EMD algorithm} 
\end{algorithm}

\section{Target Modeling Based on Histograms of Sparse Codes}

Histogram of sparse codes (HSC) has been widely
used as feature descriptors in many fields \citep{zhang2013sparse}.
Given the image set of the first $L$ image templates from a video,
a set of $K$ overlapped local image patches are sampled by a sliding
window of size $m\times n$ from each template to build a dictionary
$\mathbf{\Phi}\in\mathbb{R}^{(mn)\times(LK)}$. Each column of $\mathbf{\Phi}$
is a basis vector, which is a vectorized local image patch extracted
from the set of image templates. The basis vectors are overcomplete
where $mn<LK$. Similarly, for a given image template
target $I$, a set of overlapped local image patches $\mathbf{E}=\left\{ \epsilon_{r}|\epsilon_{r}\in\mathbb{R}^{(mn)\times1},r=1\cdots J\right\} $
are sampled by the same sliding window of size $m\times n$ with the
step size of one pixel. Each image patch $\epsilon_{r}$, which represents
one fixed part of the target object, can be encoded as a linear combination
of a few basis vectors of the dictionary $\mathbf{\Phi}$ as follows
\begin{equation}
\epsilon_{r}=\mathbf{\Phi}\mathbf{a}_{r}+\mathbf{n}
\end{equation}
where $\mathbf{a}_{r}\in\mathbb{R}^{(LK)\times1}$ is the coefficient
vector which is sparse and $\mathbf{n}\in\mathbb{R}^{(mn)\times1}$
is the noise vector. The coefficient $\mathbf{a}_{r}$ is computed
by solving the following $L_{1}$ norm minimization problem \citep{zhang2013sparse,mairal2014sparse}
\begin{equation}
\begin{array}{cc}
\underset{\mathbf{a}_{r}}{\mathbf{\mathrm{min}}} & \left\Vert \epsilon_{r}-\mathbf{\Phi}\mathbf{a}_{r}\right\Vert _{2}^{2}+\lambda\left\Vert \mathbf{a}_{r}\right\Vert _{1}\\
\mathrm{s.t.} & (\mathbf{a}_{r})_{k}\geq0,\forall k
\end{array}\label{eq:lagrange}
\end{equation}
where $\mathbf{a}_{r}=\left[\begin{array}{ccccccc}
a_{11}, & \cdots, & a_{1K}, & \cdots, & a_{L1}, & \cdots, & a_{LK}\end{array}\right]^{T}$ is the sparse coefficients of the local patch, $a_{ij}$ corresponds
to the $\mathrm{\mathit{j}\mathrm{th}}$ patch of the $i\mathrm{\mathrm{th}}$
image template of the dictionary, and $\lambda$ is the Lagrange multiplier. 

Once a solution to (\ref{eq:lagrange}) is obtained,
the maximum-alignment-pooling method is used to construct the sparse
coding histograms. Combining the coefficients corresponding to the
dictionary patches that have the same locations in the template using
$\bar{a}_{j}=\sum_{i=1}^{L}a_{ij}$ \citep{jia2016visual}, a new vector
$\bar{\mathbf{a}}_{r}=[\begin{array}{ccc}
\bar{a}_{1}, & \cdots, & \bar{a}_{j}\end{array}]^{T}\in\mathbb{R}^{K\times1}$ is formulated. The weight of the $r\mathrm{th}$ local image patch
$\epsilon_{r}$ in the histogram of sparse codes is computed by using
$\hat{p}_{ru}=\left\Vert \bar{\mathbf{a}}_{r}\right\Vert _{\infty}$.
The $\hat{p}_{ru}$ value corresponds to the $u\mathrm{th}$ image
patch from $\bar{\mathbf{a}}_{r}$. With $J$ local image patches
from the template target, the histogram is constructed as
\begin{equation}
\hat{\mathbf{p}}=[\begin{array}{c}
\hat{p}_{11},\cdots,\hat{p}_{ru},\cdots,\hat{p}_{JK}\end{array}]^{T}\in\mathbb{R}^{J\times1}\label{eq:22-1}
\end{equation}

In the spatial space, the Epanechnikov kernel is used to represent
the template. The Epanechnikov kernel \citep{comaniciu2003kernel}
is an isotropic kernel with a convex profile which assigns smaller
weights to pixels away from the center.\textcolor{black}{{} Given the
target histogram $\hat{\mathbf{p}}$ in (\ref{eq:22-1}), the isotropic
kernel is applied to generate the histograms of target weighted by
the spatial locations. The weights of the histogram of the target
$w_{T,u}$ are computed using}
\begin{equation}
w_{T,u}=\gamma\sum_{r=1}^{J}\left(1-\left\Vert \frac{\mathbf{c}_{r}}{h}\right\Vert ^{2}\right)\left|\hat{p}_{ru}\right|\label{eq:sparse target model}
\end{equation}
where $\mathbf{c}_{r}$ is the center of the $r\mathrm{th}$ image
patch of the template target, $h$ is template size
and $\gamma$ is the normalization constant. The
candidate histogram $\hat{\mathbf{q}}$ is built in the same way as
$\hat{\mathbf{p}}$. An isotropic kernel is applied to the elements
of the $\hat{\mathbf{q}}$ for generating the histogram of candidate
with spatial locations. The weights of the candidate histogram $w_{C,v}(\mathbf{x_{\mathit{i}}-\mathbf{y}})$
are computed using
\begin{equation}
w_{C,v}(\mathbf{x_{\mathit{i}}-\mathbf{y}})=\gamma\sum_{r=1}^{J}\left(1-\left\Vert \frac{\mathbf{c}_{r}-\mathbf{y}}{h}\right\Vert ^{2}\right)\left|\hat{q}_{rv}\right|\label{eq:sparse candidate model}
\end{equation}
where $\mathbf{y}$ is the displacement of the $r\mathrm{th}$ image
patch of the template candidate. The ground distance $d_{uv}$ for
the EMD in (\ref{eq:4-1}) is defined by
\begin{equation}
d_{uv}=\alpha\left\Vert \mathbf{\epsilon}_{u}-\mathbf{\epsilon}_{v}\right\Vert _{2}^{2}+(1-\alpha)\left\Vert \mathbf{c}_{u}-\mathbf{c}_{v}\right\Vert _{2}^{2}
\end{equation}
where $\alpha\in\left(0,1\right)$ is the weighting coefficient, $\mathbf{\epsilon}_{u}\in\mathbb{R}^{(mn)\times1}$,
$\mathbf{\epsilon}_{v}\in\mathbb{R}^{(mn)\times1}$ are the vectors
of the normalized pixel values of the image patch from the target
and candidate templates, sampled in the same way as the image patches
from the dictionary, and $\mathbf{c}_{u}$, $\mathbf{c}_{v}$ are
the corresponding centers of the image patches.

\section{Extensions of The Tracking Algorithm}

\subsection{Template Update}

In order to make the tracker robust to significant
appearance variations during long video sequences, the outdated templates
in the dictionary should be replaced with the recent ones. To adapt
to the appearance variations of the target and alleviate the drift
problem only the latest template in dictionary is replaced based on
the weight $\omega_{i}$, which is computed by
\begin{equation}
\omega_{i}=\gamma_{0}^{\varDelta i}\times\exp(-D_{k}^{*})\label{eq:template update}
\end{equation}
where $\omega_{i}$ is the weight associated with the template, $\gamma_{0}$
is a constant, $\varDelta i$ is the time elapsed since the dictionary
was last updated measured in terms of image index $k$ and $D_{k}^{*}$
is the EMD value corresponding to the template $I_{k}$.

If the weight of the current template based on (\ref{eq:template update})
is smaller than the weight of the latest template in the dictionary,
the template is replaced with the current one. In order to avoid the
errors and noises affecting the dictionary update algorithm, the reconstructed
template is used to replace the one in the dictionary. Firstly, the
following problem is solved in order to recompute the sparse code
coefficients, $\mathbf{a}_{k}$,
\begin{equation}
\underset{\mathbf{a}_{k}}{\mathbf{\mathrm{min}}}\left\Vert I_{k}-[\begin{array}{cc}
\mathbf{\Phi_{\mathrm{T}}} & \mathbf{I}_{mn\times mn}\end{array}]\mathbf{a}_{k}\right\Vert _{2}^{2}+\lambda\left\Vert \mathbf{a}_{k}\right\Vert _{1}\label{eq:reconstrrcut template}
\end{equation}
where $\mathbf{\Phi_{\mathrm{T}}\in\mathbb{R}}^{(mn)\times K}$ is
a dictionary formed using the vectorized template image with the size
$m\times n$ as columns, $\mathbf{I}_{mn\times mn}$ is the identity
matrix, $\mathbf{a}_{k}\in\mathbb{R}^{K+mn}$ is the vector of the
sparse coding coefficients, and $\lambda$ is the Lagrange multiplier
(cf., \citep{jia2016visual}). Then the reconstructed template is calculated
using $\mathbf{\Phi_{\mathrm{T}}}\mathbf{a}_{k}^{*}$, where $\mbox{\ensuremath{\mathbf{a}}}_{k}^{*}\in\mathbb{R}^{K}$
is computed using components of $\mathbf{a}_{k}$ corresponding to
the dictionary. The reconstructed template is used to replace the
latest template in the dictionary. The detailed steps of the update
scheme are given in Algorithm \ref{template update procedure}.
\begin{algorithm}[H]
\SetKwInOut{Input}{Input}

\SetKwInOut{Output}{Output}

\Input{The tracked template $I_{k}$ and the EMD value $D_{k}^{*}$
at frame $k$, the current dictionary $\Phi_{i-1}=\left[\begin{array}{cccc}
\mathbf{d}_{1}, & \mathbf{d}_{2}, & \cdots, & \mathbf{d}_{K}\end{array}\right]$ at index $i-1$, and the associated weights of the latest template
in dictionary $\omega_{i-1}$.}

\Output{The updated dictionary $\Phi_{i}$ and weights $\omega_{i}$.}

Compute the weight of the current template using $\omega_{i}=\exp(-D_{k}^{*})$\;

Update $\omega_{i-1}$ via (\ref{eq:template update})\;

\If{$\omega_{i}<\omega_{i-1}$ }{

$\omega_{i-1}\leftarrow\omega_{i}$\; 

Calculate the reconstructed template via (\ref{eq:reconstrrcut template})\;

$\mathbf{d}_{K}\leftarrow\mathbf{\Phi_{\mathrm{T}}}\mathbf{a}_{k}^{*}$\;
}

\protect\caption{Template update procedure.}
\label{template update procedure} 
\end{algorithm}

\subsection{Gyroscope Data Fusion for Rotation Compensation}

The general idea of the gyro-aided iEMD tracking algorithm is combining
the image frames from the camera with the angular rate generated by
the gyroscope for visual tracking. \textcolor{black}{Synchronization
of the camera and the gyroscope in time is required. The spatial relationship
between the camera and the gyroscope must also b}e pre-calibrated.
Then, the angular rate generated by the gyroscope is applied to compensate
for the ego-motion of the camera. After the compensation of the ego-motion
of the camera, the iEMD tracker is \textcolor{black}{applied for tracking.}
In this section, the gyro-aided iEMD tracking algorithm is developed
and illustrated.

When a camera is mounted on a moving robot, the motion of the camera
will cause a large displacement of the target between two consecutive
frames. If the displacement is larger than the convergence region,
the tracking algorithm will become susceptible to the large appearance
changes and fail \citep{comaniciu2003kernel,Hwangbo2011,ravichandar2014gyro}.
In order to improve the robustness of the tracking algorithm, the
displacement caused by the camera rotation is estimated and compensated
by fusing the data from the gyroscope, which is a commonly used sensor
on flying robots. The rotation of the camera causes a larger displacement
of the target compared with the translation movement in video-rate
frames. Thus, the translation is neglected here.

The gyroscope provides the angular rate along three axes, which measure
the pan, tilt, and roll of small time intervals $\varDelta t$. In
the case of pure rotation without translation, the angular rate $\omega_{y}$
is obtained along three axes $x$, $y$ and $z$. Let $q(k),\,q(k+1)\in\mathbb{H}$
denote the quaternion of two frames $k$ and $k+1$\textcolor{black}{{}
during time $\varDelta t$, the r}elationship between them is given
as (cf. \citep{spong2006robot}) 
\begin{equation}
q(k+1)=q(k)+\frac{1}{2}\Omega(\omega)\cdot q(k)\cdot\varDelta t
\end{equation}
where $\Omega(\omega)$ is the skew-symmetric matrix of $\omega$
as 
\begin{equation}
\Omega(\omega)=\left[\begin{array}{cc}
0 & -\omega^{T}\\
\omega & -[\omega]_{\times}
\end{array}\right]
\end{equation}
After the quaternion $q(k+1)=m+a\mathbf{i}+b\mathbf{j}+c\mathbf{k}$
is normalized and updated, the rotation matrix $R_{k}^{k+1}$ is calculated
as 
\begin{equation}
R_{k}^{k+1}=\left[\begin{array}{ccc}
1-2b^{2}-2c^{2} & 2ab-2cm & 2ac+2bm\\
2ab+2cm & 1-2a^{2}-2c^{2} & 2bc-2am\\
2ac-2bm & 2bc+2am & 1-2a^{2}-2b^{2}
\end{array}\right]\label{eq:22}
\end{equation}
Thus, the estimated homography matrix between two templates is estimated
by 
\begin{equation}
\mathbb{\mathcal{H}}_{\mathrm{gyro}}=KR_{k}^{k+1}K^{-1}\label{eq:gyro-hom}
\end{equation}
where, $K$ is the intrinsic camera calibration matrix that is accessed
by calibrating the camera. The homography matrix is update\textcolor{black}{d
to the newest frame location $p(k+1)=[x_{c},y_{c},1]^{T}$, where
$(x_{c},y_{c})$ is the center point of the template, based} on the
following equations: 
\begin{equation}
\mathbb{\mathcal{H}}^{k+1}=\mathbb{\mathcal{H}}^{k}\mathbb{\mathcal{\ H}}_{\mathrm{gyro}}^{-1}\label{eq:24}
\end{equation}
\begin{equation}
p(k+1)=\mathbb{\mathcal{H}}^{k+1}\cdot p(k)\label{eq:25}
\end{equation}
\textcolor{black}{for the first frame, $\mathbb{\mathcal{H}}^{0}=\mathbf{I}_{3\times3}$.
This new location $p(k+1)=[x_{c},y_{c},1]^{T}$ is then }used as the
initial guess of the object candidate and the probability of the tracking
algorithm to find the location of the object target in the new video
frame is improved.

The pseudo-code for gyro-aided iEMD algorithm is given in Algorithm
\ref{Gyro-aided iterative EMD algorithm}.

\begin{algorithm}[H]
Set the maximum iteration number ${n_{iter}}$, ${n_{scal}}$\;Capture
the image ${I_{0}}$\; \If{$t=0$}{ Display the first image ${I_{t}}$
\; Request user to select template to be tracked \;

Construct the target model from the template\;

}

\While{tracking} { Capture the image $I_{t+1}$\; Obtain the
angular rate from gyroscope\; Integrate angular rates to obtain inter-frame
rotation, $R_{t}^{t+1}$ using (\ref{eq:22})\; Compute 2D homography
$\mathbb{\mathcal{H}}_{\mathrm{gyro}}$ using (\ref{eq:gyro-hom})\;
Initialize the location of the template using (\ref{eq:25})\; Track
the target by the iEMD tracking algorithm from Algorithm 1}

\protect\protect\protect\caption{Gyro-aided iEMD tracking algorithm}
\label{Gyro-aided iterative EMD algorithm} 
\end{algorithm}

\section{Experiments }

In this section, the iEMD algorithm is validated on real datasets.
The algorithm is implemented by M\textcolor{black}{ATLAB} R2015b,
the C code in \citep{rubner2000earth} is adopted for the EMD calculation,
and the software in \citep{mairal2014sparse} is used for sparse modeling.
The platform is Microsoft Windows 7 professional with Intel(R) Core(TM)
i5-4590 CPU. Eight publicly available datasets are chosen to validate
the iEMD tracking algorithm. The main attributes of the video sequences
are summarized in Table \ref{tab:The-main-attributes of datasets}.
The Car2, Walking, Woman, Subway, Bolt2, Car4, Human8, and Walking2
sequences are from the visual tracker benchmark \citep{WuLimYang13}
(CVPR 2013, \url{http://www.visual-tracking.net}). The length of
the sequences varies between 128 and 913 frames with one object being
tracked in each frame. 

The tracker is initialized with the ground-truth bounding box of the
target in the first frame. Then the tracking algorithm runs till the
end of the sequence and generates a series of the tracked bounding
boxes. Tracking results from consecutive frames are compared with
the ground truth bounding boxes provided by this dataset. The relative
overlap measure is used to evaluate this algorithm as \citep{WuLimYang13}
\begin{equation}
\mathrm{overlap}=\frac{\mathbf{R}_{tr}\cap\mathbf{R}_{gt}}{\mathbf{R}_{tr}\cup\mathbf{R}_{gt}}
\end{equation}
where $\mathbf{R}_{tr}$ is the tracking result, represented by the
estimated image region occupied by the tracked object, $\mathbf{R}_{gt}$
is the ground truth bounding box. $\mathbf{R}_{tr}\cap\mathbf{R}_{gt}$
is the intersection and $\mathbf{R}_{tr}\cup\mathbf{R}_{gt}$ is the
union of the two regions. The range of the relative overlap is from
$0$ to $100\%$ .

\subsection{Results for the iEMD tracker with sparse coding histograms}

In this subsection, the performance of the iEMD
tracker with sparse coding histograms and the template update method
is evaluated using the eight sequences. In our approach, the object
windows are re-sized to $32\times32$ pixels for all the sequences,
except for the Walking sequence, in which the object windows are resized
to $64\times32$ pixels due to the smaller object size. The local
patches in each object window are sampled with the size $16\times16$
pixels with step size $8$ in sequences like Car4, Walking and Car2.
For other sequences, the local patches in each object window are sampled
with the size $8\times8$ pixels with step size $4$. In the case
of the abrupt motions of the object, $4$ more particles are generated
by moving the template in the surrounding area of the initial object
position. For each particle, the template is enlarged and shrunk by
$2\%$ in case of the scale variations. 

\begin{table}
\begin{centering}
\protect\caption{The main attributes of the video sequences. Target size: the initial
target size in the first frame, IV: illumination variation, SV: scale
variation, OCC: occlusion, DEF: deformation, MB: motion blur, FM:
fast motion, BC: background clutters.\label{tab:The-main-attributes of datasets}}

\par\end{centering}

\begin{centering}
\begin{tabular}{ccccccccccc}
\hline 
Sequence & Frames & Image size & Target size & IV & SV & OCC & DEF & MB & FM & BC\tabularnewline
\hline 
\hline 
Car4 & $659$ & $360\times240$ & $107\times87$ & $\checked$ & $\checked$ &  &  &  &  & \tabularnewline
\hline 
Walking & $412$ & $768\times576$ & $24\times79$ &  & $\checked$ & $\checked$ & $\checked$ &  &  & \tabularnewline
\hline 
Woman & $550$ & $352\times288$ & $21\times95$ & $\checked$ & $\checked$ & $\checked$ & $\checked$ & $\checked$ & $\checked$ & \tabularnewline
\hline 
Subway & $175$ & $352\times288$ & $19\times51$ &  &  & $\checked$ & $\checked$ &  &  & $\checked$\tabularnewline
\hline 
Bolt2 & $293$ & $480\times270$ & $34\times64$ &  &  &  & $\checked$ &  &  & $\checked$\tabularnewline
\hline 
Car2 & $913$ & $320\times240$ & $64\times52$ & $\checked$ & $\checked$ &  &  & $\checked$ & $\checked$ & $\checked$\tabularnewline
\hline 
Human8 & $128$ & $320\times240$ & $30\times91$ & $\checked$ & $\checked$ &  & $\checked$ &  &  & \tabularnewline
\hline 
Walking2 & $500$ & $384\times288$ & $31\times115$ &  & $\checked$ & $\checked$ &  &  &  & \tabularnewline
\hline 
\end{tabular}
\par\end{centering}

\end{table}

\begin{figure}[h!]
\begin{center}
\includegraphics[width=18cm]{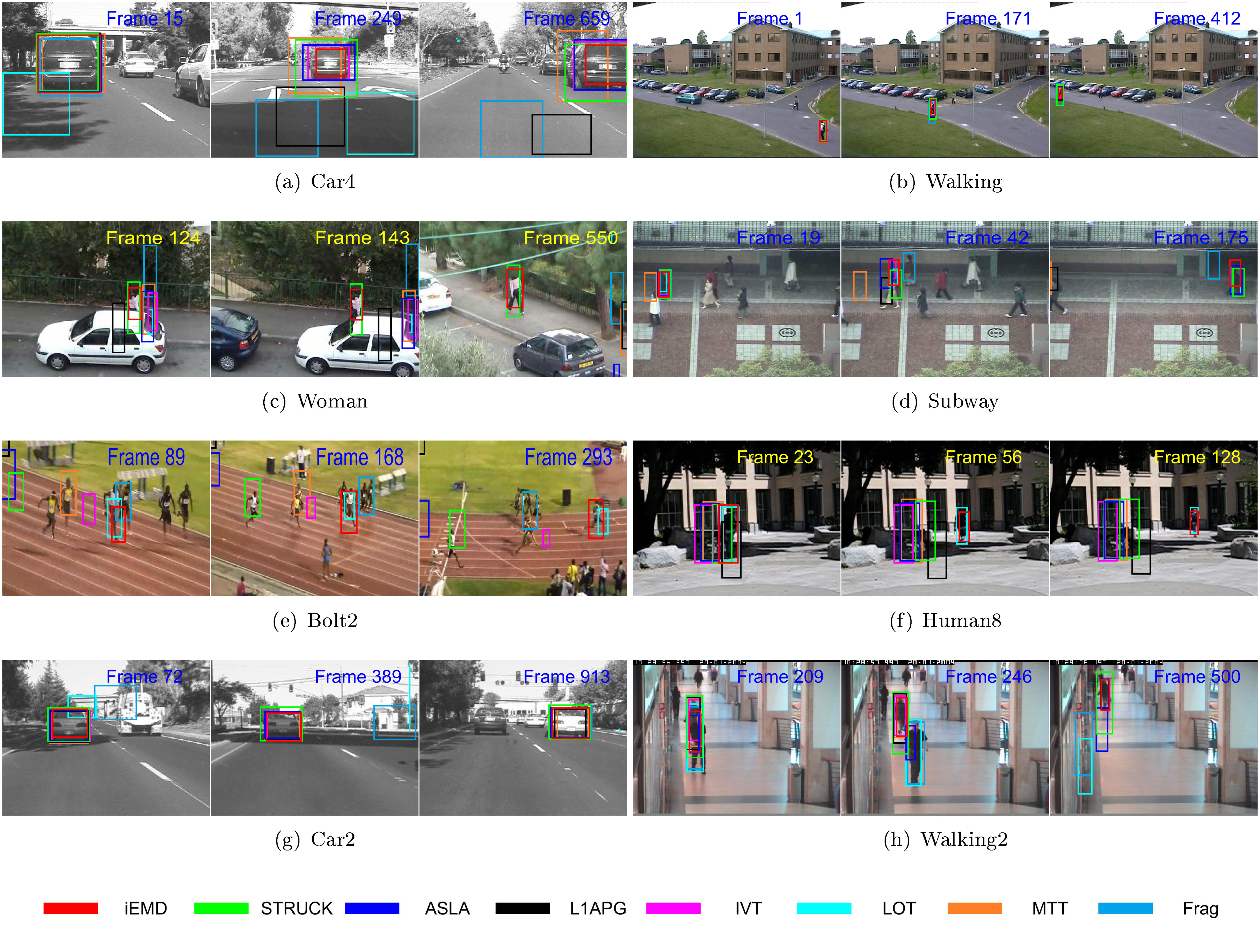}
\end{center}
\caption{ The visual tracking results obtained by the eight tracking algorithms on the eight video sequences.}\label{fig:The-visual-tracking}
\end{figure}

\begin{figure}[h!]
\begin{center}
\includegraphics[width=18cm]{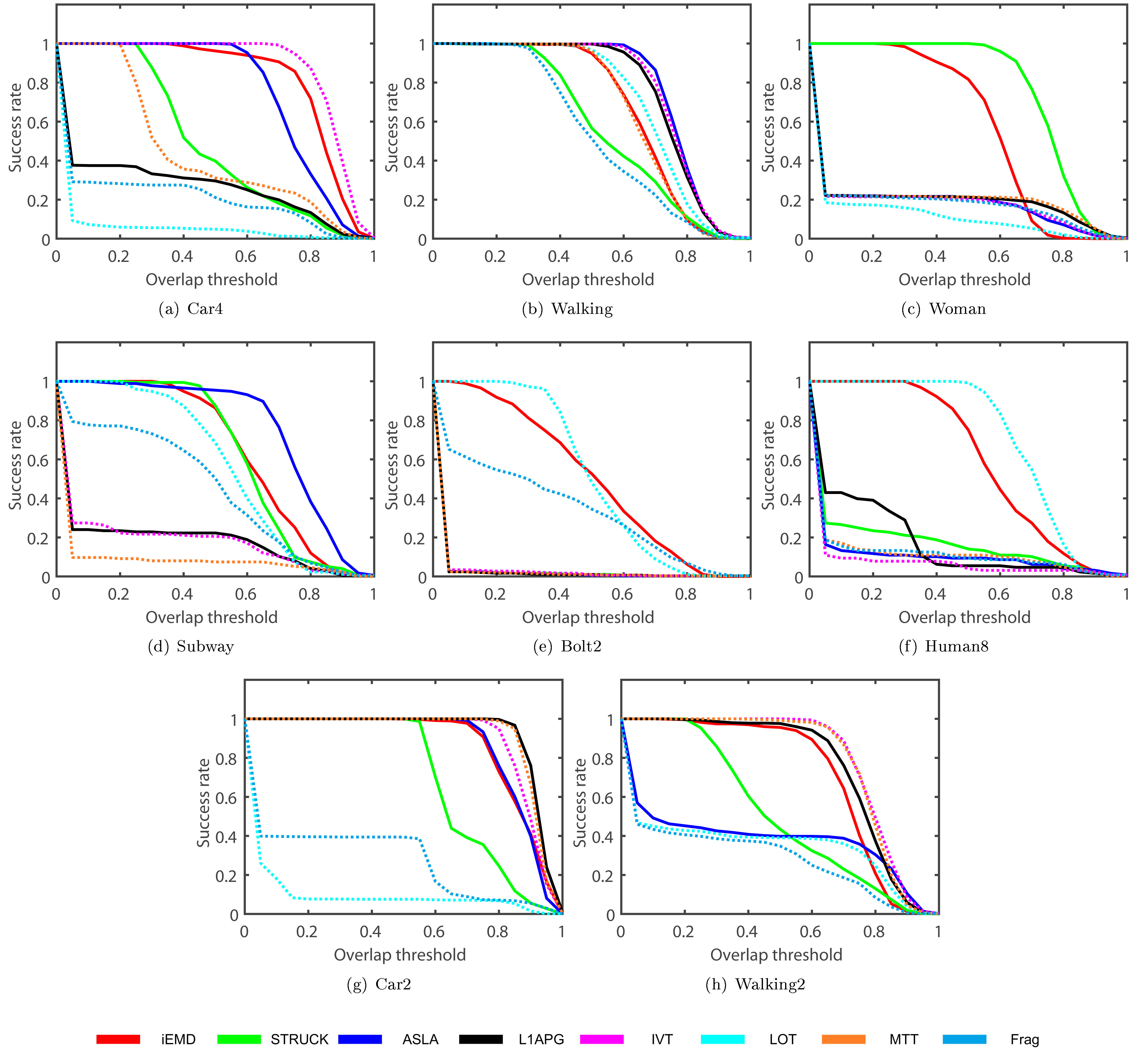}
\end{center}
\caption{ Success plots ((a)-(j)) for the eight tracking algorithms on the eight sequences.}\label{fig:fig:Sucess-plots}
\end{figure}

The performance of the proposed algorithm is compared
with seven state-of-the-art tracking algorithms on eight
video sequences. These state-of-the-art trackers include: ASLA \citep{jia2012visual},
Frag \citep{adam2006robust}, IVT \citep{ross2008incremental}, L1APG
\citep{mei2011robust}, LOT \citep{oron2012locally}, MTT \citep{zhang2012robust},
and STRUCK \citep{hare2011struck}. The source codes of the trackers
are downloaded from the corresponding web pages and the default parameters
are used. The average percentage overlap obtained by all the tracking
algorithms on eight video sequences are reported in Table \ref{tab:The-average-overlap}.
The iEMD tracker achieves the highest average overlap over all the
sequences. The iEMD tracker also achieves the second best tracking
results on the 5 out of $8$ sequences. In Fig. \ref{fig:Sucess-plots,
the success plots obtained by all the tracking algorithms on }eight
video sequences are shown. The success plot shows the ratios of frames
at the different thresholds of the relative overlap values varied
from $0$ to $1$.

\begin{center}
\begin{table}[h]
\begin{centering}
\protect\caption{The average overlap (in percentage) obtained by the tracking algorithms
on eight datasets. For each sequence, the first, second and third
ranks are marked in red, green and blue respectively. The last row
is the average value of the percentage overlap for each tracker over
all sequences.\label{tab:The-average-overlap}}

\par\end{centering}

\begin{centering}
\begin{tabular}{ccccccccc}
\hline 
Sequence & ASLA & Frag & IVT & L1APG & LOT & MTT & STRUCK & iEMD\tabularnewline
\hline 
\hline 
Car4 & \textcolor{blue}{75.4} & 18.8 & \textcolor{red}{87.6} & 24.9 & 4.2 & 44.7 & 48.9 & \textcolor{green}{82.0}\tabularnewline
\hline 
Walking & \textcolor{red}{77.2} & 53.7 & \textcolor{green}{76.6} & \textcolor{blue}{75.3} & 70.4 & 66.6 & 57.1 & 67.1\tabularnewline
\hline 
Woman & 14.8 & 14.7 & 14.7 & 16.2 & 8.9 & \textcolor{blue}{16.7} & \textcolor{red}{73.2} & \textcolor{green}{60.7}\tabularnewline
\hline 
Subway & \textcolor{red}{75.6} & 44.0 & 15.9 & 16.2 & 56.0 & 6.8 & \textcolor{blue}{62.6} & \textcolor{green}{63.9}\tabularnewline
\hline 
Bolt2 & 1.1 & \textcolor{blue}{32.6} & 1.6 & 1.1 & \textcolor{red}{51.8} & 1.1 & 1.2 & \textcolor{green}{50.1}\tabularnewline
\hline 
Car2 & 86.4 & 25.9 & \textcolor{blue}{89.3} & \textcolor{red}{92.4} & 8.6 & \textcolor{green}{91.5} & 68.8 & 86.2\tabularnewline
\hline 
Human8 & 8.8 & 9.7 & 5.5 & \textcolor{blue}{15.6} & \textcolor{red}{70.4} & 9.8 & 14.7 & \textcolor{green}{60.2}\tabularnewline
\hline 
Walking2 & 37.1 & 27.4 & \textcolor{red}{79.5} & \textcolor{blue}{75.6} & 33.5 & \textcolor{green}{78.5} & 51.0 & 71.2\tabularnewline
\hline 
\hline 
Average & \textcolor{green}{47.1} & 28.4 & 46.3 & 39.7 & 38.0 & 39.5 & \textcolor{blue}{47.2} & \textcolor{red}{67.7}\tabularnewline
\hline 
\end{tabular}
\par\end{centering}

\end{table}

\par\end{center}

Representative tracking results obtained by iEMD
algorithm are shown in Fig. \ref{fig:The-visual-tracking}. In the
Human8 and Bolt2 sequences, the targets have significant illumination
variations, and deformations, respectively. Only LOT and iEMD trackers
are able to track the target in all the frames. Both of them use the
EMD as the similarity measure and their appearance models are based
on local image patches, which make the trackers more robust to illumination
changes and deformations \citep{oron2012locally,rubner2000earth}.
In woman sequence, all the trackers start to drift away from the target
in frame $124$ except for the iEMD and STRUCK trackers. For the Car2
and Car4 sequences, there are significant illumination changes when
the targets pass underneath the trees and the overpasses. The LOT
and Frag trackers start drifting away from frame $72$ in Car2 sequences.
In Car4 sequence, the LOT tracker starts to lose the target from frame
$15$, and the Frag and L1APG trackers drift away when the car passes
the overpass in frame $249$. In Walking2 sequence, the LOT, Frag,
and ASLA trackers start tracking the wrong target in frame $246$,
due to the similar colors of the clothes between the two people.

\subsection{Results for the Gyro-aided iEMD tracking algorithm }

The test of the gyro-aided iEMD tracking algorithm is conducted using
the sequence including 100 frames from the dataset provided by CMU
\citep{Hwangbo2011}. The size of the template is changed by $\pm10\%$
and the template with the best scale is found, giving the smallest
$EMD$. The images are taken in front of a desk with motions, such
as shaking and rotation. The frame sequences have a resolution of
$640\times480$ at $30\mathrm{FPS}$. The gyroscope is carefully aligned
with the camera and the tri-axial gyroscopic values are sampled at
$11\mathrm{Hz}$ in the range of $\pm200\mathrm{deg/sec}$ \citep{Hwangbo2011}.
Using the time stamps of the camera and the gyroscope, the angular
rate data are synchronized with the frames captured by the camera.

\begin{figure}[H]
\centering{}\centering{}\includegraphics[width=0.7\columnwidth]{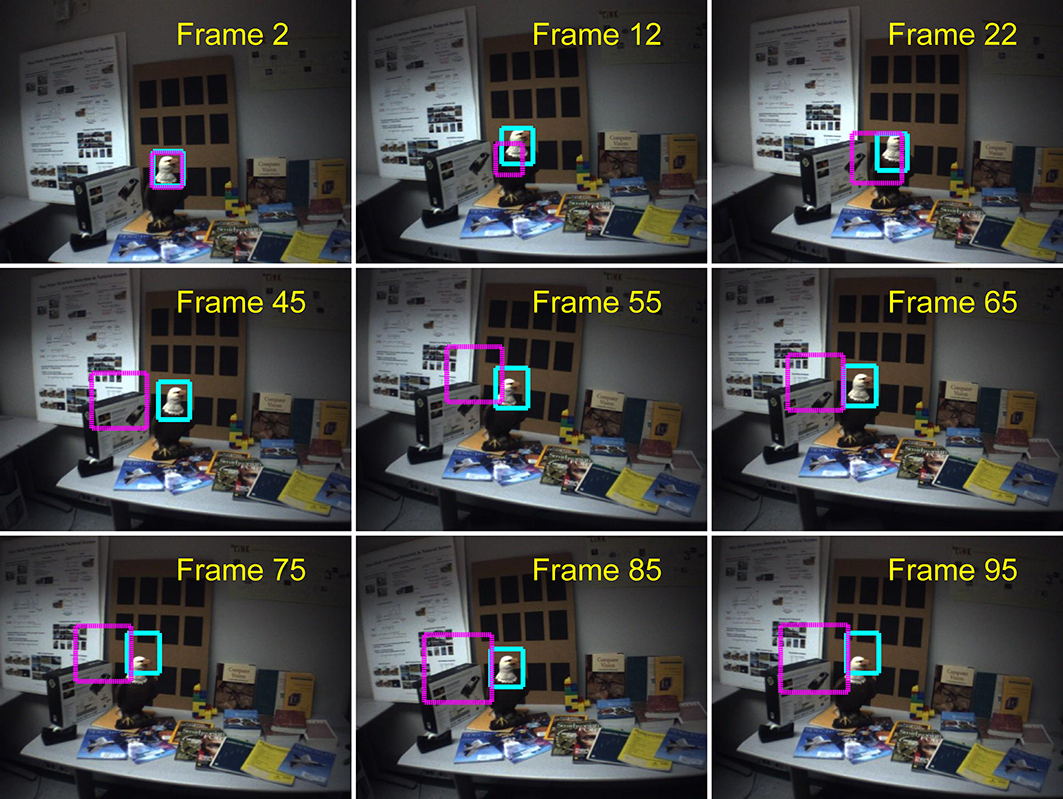}\protect\protect\caption{Results of the iEMD tracker in presence of rapid camera motion; the
magenta boxes indicate the results of the iEMD tracker without the
gyroscope information, and the cyan boxes indicate the results of
the gyro-aided iEMD tracker.}
\label{fig.1} 
\end{figure}

The comparisons between the tracking results using the iEMD tracker
with and without the gyroscope information are illustrated in Fig.
\ref{fig.1}. The head of the eagle is chosen as the target and the
ground truth is manually labeled in each frame. The magenta box indicates
the estimated image region without using the gyroscope data, and the
cyan box is the tracking results of the gyro-aided iEMD tracker. Without
the gyroscope data, the tracker loses the target after the frame $25$.
However, the head of the eagle is successfully tracked with our gyro-aided
iEMD tracking algorithm.

The performances of the iEMD tracker with and without the gyroscope
information on the CMU sequence are summarized in Table \ref{tab:Evaluationimu}.
The value of the average overlap, the percentage of the total frame
numbers of which the overlap is greater than $0$ and $40\%$ are
reported. Gyroscope information provides a good initial position for
the iEMD tracker to estimate the location of the target. Thus, the
gyro-aided iEMD tracking algorithm is robust to the rapid movements
of the camera.

\begin{table}[H]
\protect\caption{\label{tab:Evaluationimu}Evaluation results on the CMU dataset using
the iEMD tracker with and without the gyroscope information.}

\begin{centering}
\begin{tabular}{ccc}
\hline 
\textcolor{black}{Relative overlap with the ground truth } & \textcolor{black}{Gyro-aided } & \textcolor{black}{No gyro-aided}\tabularnewline
\hline 
\hline 
\textcolor{black}{Average overlap (\%) } & \textcolor{black}{$50.7$ } & \textcolor{black}{$11.9$ }\tabularnewline
\hline 
\textcolor{black}{Overlap $>0$ } & \textcolor{black}{$100$ } & \textcolor{black}{$70.7$ }\tabularnewline
\hline 
\textcolor{black}{Overlap $>40\%$ } & \textcolor{black}{$66.7$ } & \textcolor{black}{$9.1$ }\tabularnewline
\hline 
\end{tabular}
\par\end{centering}

\end{table}

\subsection{Discussion}

As a cross-bin metric for the comparison of the
histograms, the advantages of the EMD are demonstrated in situations
such as illumination variation, object deformation and partial occlusion.
The iEMD algorithm uses the transportation-simplex algorithm for calculating
the EMD in the experiments, of which the practical running time complexity
is the supercubic (a complexity in $\Omega(N^{3})\bigcap O(N^{4})$)
\citep{rubner2000earth}, where $N$ represents the number of the histogram
bins. Other algorithms for calculating the EMD can be used to further
shorten the running time \citep{ling2007efficient,pele2009fast}. The
experimental results, especially the Human8 and Bolt2 sequences, show
that the iEMD tracker is robust to the appearance variations. The
experimental results of Walking2 shows that the iEMD tracker can discriminate
the target from the surroundings with similar colors. The tracking
results from Woman and Subway sequences demonstrate the robustness
to partial occlusions. Since the local sparse representation is adopted
as the appearance model, the methods such as the trivial templates,
learning dictionary from the target and background images, could be
adopted to improve the performance of iEMD tracker. As a gradient
descent based dynamic model, the iEMD tracker, which provides good
location prediction, can be further improved with more effective particle
filters. The metrics used by sparse coding, such as the largest sum
of the sparse coefficients or the smallest reconstruction error, can
be combined with the EMD to make the tracker more discriminant. 

\section{Conclusion}

This paper presents iEMD and gyro-aided iEMD visual
tracking algorithms. The local sparse representation is used as the
appearance model for the iEMD tracker. The maximum-alignment-pooling
method is used for constructing a sparse coding histogram which reduces
the computational complexity of the EMD optimization. The template
update algorithm based on the EMD is also presented. The iEMD tracker
is robust to variations in appearance of the target, deformations
and partial occlusions. Experiments conducted on eight publicly available
datasets show that the iEMD tracker is robust to the illumination
changes, deformations and partial occlusions of the target. To validate
the gyro-aided iEMD tracking algorithm, experimental results from
the CMU dataset, which contains rapid camera motion are presented.
Without the gyroscope measurements, the iEMD tracker fails on the
CMU dataset. With the help of the gyroscope measurements, the iEMD
algorithm is able to lock onto the target and track it successfully.
The above experimental results show that the proposed iEMD tracking
algorithm is robust to the appearance changes of the target as well
as the ego-motion of the camera. 

\section*{Acknowledgment}

The authors would like to thank Prof. Peter Willett, Iman Salehi and
Harish Ravichandar for their help.

\bibliographystyle{frontiersinSCNS_ENG_HUMS}
\bibliography{revisedpaper}

\end{document}